\documentclass[conference]{IEEEtran}
\IEEEoverridecommandlockouts
\usepackage{hyperref}
\usepackage{amsmath,amssymb,amsfonts}
\usepackage{amssymb}
\usepackage{algorithmic}
\usepackage{graphicx}
\usepackage{textcomp}
\usepackage{xcolor}
\usepackage{afterpage}
\usepackage{mathtools}
\usepackage{amsmath}
\usepackage{breqn}
\usepackage{booktabs}
\usepackage[noadjust]{cite}

\usepackage{tikz}
\usepackage{circuitikz}
\usepackage{pgfplots}
\pgfplotsset{compat=1.14}

\definecolor{Set1-7-1}{RGB}{228,26,28}
\definecolor{Set1-7-2}{RGB}{55,126,184}
\definecolor{Set1-7-3}{RGB}{77,175,74}
\definecolor{Set1-7-4}{RGB}{152,78,163}
\definecolor{Set1-7-5}{RGB}{255,127,0}
\definecolor{Set1-7-6}{RGB}{166,86,40}
\definecolor{Set1-7-7}{RGB}{0,0,0}

\def\BibTeX{{\rm B\kern-.05em{\sc i\kern-.025em b}\kern-.08em
    T\kern-.1667em\lower.7ex\hbox{E}\kern-.125emX}}

\begin{document}

\title{Non-Linear Self-Interference Cancellation via \\Tensor Completion}

\author{\IEEEauthorblockN{Freek Jochems and Alexios Balatsoukas-Stimming}
\IEEEauthorblockA{Department of Electrical Engineering, Eindhoven University of Technology, The Netherlands\\f.m.j.jochems@student.tue.nl, a.k.balatsoukas.stimming@tue.nl}} 

\maketitle

\begin{abstract}
Non-linear self-interference (SI) cancellation constitutes a fundamental problem in full-duplex communications, which is typically tackled using either polynomial models or neural networks. In this work, we explore the applicability of a recently proposed method based on low-rank tensor completion, called \emph{canonical system identification} (CSID), to non-linear SI cancellation. Our results show that CSID is very effective in modeling and cancelling the non-linear SI signal and can have lower computational complexity than existing methods, albeit at the cost of increased memory requirements.
\end{abstract}

\section{Introduction}

Full-duplex (FD) wireless systems allow bi-directional communication on the same frequency band and at the same time \cite{Jain2011, Duarte2012, Bharadia2013}. Compared to half-duplex systems, FD systems have the potential to double the transmission capacity, among many other potential advantages. However, the main challenge in FD communications is the presence of a self-interference (SI) signal, which is significantly stronger than the desired signal and needs to be canceled.

SI cancellation can be performed in both the digital domain and the radio frequency (RF) domain. Cancellation in both domains is typically required to cancel the SI signal to the level of the receiver noise floor. RF cancellation can be split in two types of cancellation: passive and active cancellation. Passive RF cancellation can be achieved through physical isolation between the transmitting and the receiving antenna. There are a number of methods to achieve this, such as beamforming, directional antennas, circulators, polarization or shielding~\cite{Everett2014}. Active RF cancellation can be performed through injecting a cancellation signal, which is done right after the receiving antenna. One of the ways this can be implemented is by coupling into the transmitted RF signal, adding a time delay, a phase rotation, and an attenuation and adding the resulting SI cancellation signal to the received SI signal \cite{Jain2011, Bharadia2013}.

\begin{figure*}[t]
  \centering
  \scalebox{0.95}{\begin{circuitikz}[scale=1]

	%%% Transmitter %%%	
	% Nodes
	\draw (4,0) node[mixer,scale=0.5] (txmixer) {};
	\draw (11,0) node[antenna,scale=0.6] (txantenna) {};
	\draw (0,-1.4) node[draw] (digcanc) {Digital Cancellation};
	
	% Connections	
	\draw (-1.5,0) to[short,*-] (0.25,0);
	\draw (0.25,0) to[short,-] ++(0,0) to[twoport,>,t=DAC] (txmixer.west) node[inputarrow]{};
	\draw (txmixer.east) to[short,-] ++(0.5,0) to[amp,>,t=\small{PA}] ++(2.25,0) to[bandpass,>,l^=BP Filter] ++(1.5,0) to ++(1.5,0) to (txantenna);
	
	% Text labels
	\draw (-1.5,0) node[left] {\small $x$};
	\draw (txmixer)+(0,0.5) node[above] {\small{IQ Mixer}};
	\draw (txmixer)+(-0.9,0) node[below] {\small $x_{\text{DAC}}$};
	\draw (txmixer)+(0.9,0) node[below] {\small $x_{\text{IQ}}$};
	\draw (txantenna)+(-4.2,0) node[below] {\small $x_{\text{PA}}$};
	
	%%% Receiver %%%
	% Nodes
	\draw (11,-3) node[antenna,scale=0.6] (rxantenna) {};
	\draw (9.5,-3) node[adder,scale=0.5] (rxadder) {};
	\draw (4,-3) node[mixer,scale=0.5] (rxmixer) {};
	\draw (txantenna)++(-1.5,-1.4) node[draw] (rfcanc) {RF Cancellation};
	\draw (0,-3) node[adder,scale=0.5] (rxadderdig) {};
	
	% Connections	
	% Digital canceler
	\draw (0,0) to (digcanc.north) node[inputarrow,rotate=-90]{};
	\draw (digcanc.south) to (rxadderdig.north) node[inputarrow,rotate=-90]{};
	% RF canceler
	\draw (txantenna)++(-1.5,0) to (rfcanc.north) node[inputarrow,rotate=-90]{};
	\draw (rfcanc.south) to (rxadder.north) node[inputarrow,rotate=-90]{};
	\draw (rxantenna) to[short] (rxadder.east) node[inputarrow,rotate=180]{};
	% Rest of chain
	\draw (rxadder.west) to ++(0,0) to[bandpass,>,l=BP Filter] ++(-2.7,0) to[amp,>,t={\rotatebox[origin=c]{180}{\small{LNA}}}] ++(-1.5,0) to[short,-] (rxmixer.east) node[inputarrow,rotate=180]{};
	\draw (rxmixer.west) to[twoport,>,t=ADC] (rxadderdig.east);
	\draw (rxadderdig.west) to[short,-*] (-1.5,-3);
	\draw [->] (txantenna)+(0.4,1.3) to[thick, out=-50, in=50, edge node={node [right] {$h_{\text{SI}}$}}]  ($(rxantenna) + (0.4,1.35)$);
	
	% Text labels
	\draw (rxantenna)+(-4.1,0) node[above] {\small $y_{\text{RX}}$};
	\draw (rxmixer)+(0.9,0) node[above] {\small $y_{\text{LNA}}$};
	\draw (rxmixer)+(-0.9,0) node[above] {\small $y_{\text{IQ}}$};
	\draw (-1.5,-3) node[left] {\small $y$};
	\draw (rxmixer)+(0,-0.55) node[below] {\small{IQ Mixer}};
	
	%%% Reference clock %%%
	\draw (4.25,-1.5) node[oscillator,scale=0.5] (ref) {};
	\draw (ref.south) to[short,-] (rxmixer.north) node[inputarrow,rotate=270]{};
	\draw (ref.north) to[short,-] (txmixer.south) node[inputarrow,rotate=90]{};
	\draw (ref)+(+0.1,0) node[right,text width=1.2cm,align=center] {\small{Local\\Oscillator}};
	
\end{circuitikz}}
  \caption{Block diagram of a FD transceiver with active RF SI cancellation and digital SI cancellation~\cite{Kurzo2020}. A few components have been omitted for simplicity, a more detailed diagram can be found in~\cite{Korpi2017}.}
  \label{fig:system}
\end{figure*}
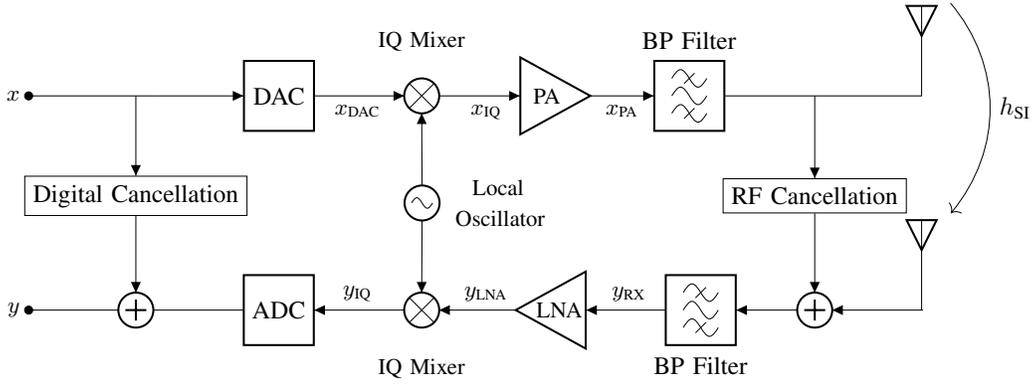

Perfect RF cancellation is difficult to achieve in practice due to complexity, cost, and various transceiver non-linearities and impairments. Thus, in general, after RF cancellation there is still a residual SI signal present. In the digital domain this signal can be canceled more easily, because conventional digital signal processing (DSP) methods can be used. However, due to the presence of non-linear components between the transmitter and receiver, non-linear DSP techniques have to be used. Fig.~\ref{fig:system} shows an overview of a FD system, which includes a digital-to-analog converter (DAC), an analog-to-digital converter (ADC), an IQ mixer, a power amplifier (PA) and a low-noise amplifier (LNA). All of these components introduce non-linear effects into the SI signal, meaning that high-complexity non-linear cancellation methods are required. Traditional non-linear methods, which have been widely used in the FD literature, rely on polynomial models \cite{Balatsoukas2015,Korpi2017,Campo2018}. More recently, machine learning techniques have also been used to model transceiver non-linearities, mostly focusing on the use of black-box and model-based neural networks (NNs)~\cite{Balatsoukas2018,Kurzo2018,Guo2018,Kristensen2019,Shi2019,Wang2019,Kurzo2020,Elsayed2020,Chen2020,Kristensen2020}, but also on other techniques such as support vector machines~\cite{Auer2020} and tree-based algorithms~\cite{Dikmese2019}.

\emph{Contribution:} Canonical system identification (CSID) is a recent machine learning method that can be used to effectively model non-linear systems using a low-rank tensor decomposition~\cite{Kargas2020}. In this work we perform an initial exploration of the applicability of CSID to the problem of non-linear SI cancellation in FD systems. To this end, we compare the SI cancellation performance and the computational and memory complexities of a CSID-based SI canceller with existing polynomial and NN-based SI cancellers. We note that the described method is also applicable to other communications-related problems that employ non-linear models, such as digital pre-distortion and mitigation of passive intermodulation in simultaneous transmit-receive systems.

\section{Self-Interference Cancellation in Full-Duplex Systems}\label{sec:background}

\subsection{Linear Self-Interference Cancellation}
For simplicity, let us assume that there is no signal-of-interest present from a remote node and only the SI signal and thermal noise is received. The objective of digital SI cancellation is to make an estimate of the complex-valued baseband SI signal $y_{SI}[n]$ based on the transmitted complex-valued baseband signal $x[n]$. This estimate, denoted by $\hat{y}_{SI}[n]$, is then substracted from the received signal $y[n]$ to cancel the SI signal. The SI cancellation performance is typically evaluated as
\begin{align}
  C_{\text{dB}} &= 10 \log_{10} \left( \frac{\sum_n |y[n]|^2}{\sum_n |y[n] - \hat{y}_{\text{SI}}[n]|^2} \right).
\end{align}
The most basic form of digital cancellation only considers linear effects and is constructed as
\begin{equation}
    \hat{y}_{\text{SI,lin}}[n]=\sum_{l=0}^{L-1}\hat{h}[l]x[n-l], \label{eq:canclin}
\end{equation}
where the channel $\hat{h}[l]$, models the SI channel $h_{\text{SI}}$ and any other memory effect in the transceiver chain and is estimated from training data using, e.g., least squares (LS) estimation.

\subsection{Non-Linear Self-Interference Cancellation}
Due to the non-linear effects introduced by components in the transceiver chain, linear cancellation is often not sufficiently powerful to model and cancel the SI, so that non-linear models  have to be used. The following widely linear memory polynomial model is commonly used in the literature~\cite{Korpi2017}
\begin{align}
	\hat{y}_{\text{SI}}[n]	& = \sum _{\substack{p=1,\\p \text{ odd}}}^P \sum_{q=0}^p\sum_{l=0}^{L-1}\hat{h}_{p,q}[l] x[n-l]^{q}x^*[n-l]^{p-q}, \label{eq:cancnonlin}
\end{align}
where the channels $\hat{h}_{p,q}[l]$ are estimated from training data using, e.g., least squares (LS) estimation.

Alternatively, both black-box NNs and model-based NNs have been shown to be an effective and low-complexity method to model and to cancel the SI signal~\cite{Balatsoukas2018,Kurzo2018,Guo2018,Kristensen2019,Shi2019,Wang2019,Kurzo2020,Elsayed2020,Chen2020,Kristensen2020}. These NNs are trained using training data and backpropagation.

\section{Non-linear Self-interference Cancellation via Tensor Completion}\label{sec : method}

\subsection{Canonical System Identification (CSID)}\label{sub : CSID}
The main idea behind CSID is that any non-linear system with $N$ inputs and a single real-valued output can be represented as an $N$-dimensional tensor $\mathcal{X}$. In particular, let the input of the non-linear system be denoted by $\mathbf{i} = \begin{bmatrix}  i_1 & \hdots & i_{N} \end{bmatrix}$, where $i_1 \in \{1,\hdots,I_n\}$ and $I_n$ denotes the number of different values $i_n$ can take on. Moreover, let the output of the system be denoted by $y$. Then, for an appropriately defined tensor $\mathcal{X}$, we have
\begin{align}
	y = \mathcal{X}\left(i_1,....,i_{N}\right). \label{eq:nltensor}
\end{align}
In practice, and in particular in communications systems, the inputs are often not integer-valued. In this case,  $i_n$ can be thought of as the label corresponding to a quantized version of the $n$-th input with $I_n$ quantization levels or, equivalently, $\log_2 I_n$ quantization bits. Without any assumptions on $\mathcal{X}$, the representation in \eqref{eq:nltensor} is not particularly relevant from a practical perspective. This is because the number of elements in $\mathcal{X}$, denoted by $|\mathcal{X}|$, is given by
\begin{align}
	|\mathcal{X}| & = \prod _{n=1}^N I_n, 
\end{align}
which can become very large even when $N$ is small (i.e., even when coarse quantization is used). For example, for $N=4$ and using $\log_2 I_n = 8$ quantization bits for all $n \in \{1,\hdots,N\}$, we have $|\mathcal{X}| = 2^{32}$.

This issue can be alleviated by imposing a low-rank constraint on the tensor $\mathcal{X}$. In particular, a tensor of rank $F$ can be represented as
\begin{align}
	\mathcal{X}\left(i_1,....,i_N\right) = \sum_{f=1}^{F}\prod_{n=1}^{N}\mathbf{A}_n(i_n,f), \label{eq:tensor}
\end{align}
where $\mathbf{A}_n$ are called the \emph{factor matrices}. This results in a significantly more compact representation of $\mathcal{X}$. Let us denote the set of $M$ inputs and outputs that are used for training by $\mathbf{x}_m,~m \in \{1,\hdots,M\},$ and $y_m,~m \in \{1,\hdots,M\}$, respectively. Then, CSID amounts to solving the following optimization problem~\cite{Kargas2020}
\begin{align}
	\min _{\mathcal{X},\{\mathbf{A}_n\}_{n=1}^N} & \quad \frac{1}{M}\sum _{m=1}^M \left|y_m - \mathcal{X}(x_m[0],\hdots,x_m[N{-}1])\right|^2 \label{eq:csid} \\
	\text{s.t.} &  \quad \mathcal{X} = \sum_{f=1}^{F}\mathbf{A}_1(:,f)\odot \hdots \odot \mathbf{A}_N(:,f) \nonumber
\end{align}
The terms $\sum_{n=1}^N \rho ||\mathbf{A}_n||^2_F$ and $\sum_{n=1}^N \mu_n ||\mathbf{T}_n\mathbf{A}_n||^2_F$ can be added to the cost function in~\eqref{eq:csid} for regularization and to add a smoothness constraint, respectively.\footnote{For more details and for the definition of $\mathbf{T}_n$, see~\cite[Section 3.1]{Kargas2020}.} The optimization problem in \eqref{eq:csid} can be solved efficiently using an alternating least squares algorithm (ALS), as described in~\cite{Kargas2020}.

We note that, the ALS algorithm in~\cite{Kargas2020} was described for a real-valued tensor $\mathcal{X}$. However, the approach can be extended to complex-valued $\mathcal{X}$ in a straightforward manner by replacing all transpose operations in \cite[Eq. (11)]{Kargas2020} with Hermitian transpose operations to obtain
\begin{align}
    \mathbf{a}^{k}_{i}  & = (\mathbf{Q}^{H}_{k}\text{diag}(\mathbf{w}_i)^{2}\mathbf{Q}_{k}+(\rho+2\mu_{k})\mathbf{I})^{-1} \nonumber \\
						& \qquad (\mathbf{Q}_{k}^{H}\text{diag}(\mathbf{w}_i)^2\mathbf{y}^{k}_{i}-\mu_{k}(\mathbf{a}_{i-1}^{k}+\mathbf{a}^{k}_{i+1})).
\end{align}
A complex-valued representation is useful in communications systems, where DSP operations are typically carried out on the complex-valued baseband signals.

\subsection{CSID-Based Non-Linear SI Cancellation}
Similarly to the NN-based cancellers of~\cite{Kurzo2020}, CSID-based SI cancellation is split into two steps. First, we use linear cancellation to remove the linear part of the SI, which significantly reduces the dynamic range of the SI signal. Then, we use CSID to model the non-linear part of the SI. More specifically, the expected outputs $y_m$ in~\eqref{eq:csid} that we use to train CSID are
\begin{align}
	y_m= y[m] - \hat{y}_{\text{SI,lin}}[m], \quad \forall m \in \mathcal{M}_{\text{train}},
\end{align}
where $\mathcal{M}_{\text{train}}$ denotes the set of training indices. As can be seen from~\eqref{eq:canclin} and \eqref{eq:cancnonlin}, a SI sample at time instant $m$ is a function of $x[m-l],~l \in \{0,\hdots,L-1\}$. Moreover, $x[m]$ are complex-valued and cannot be directly used as tensor indices (even after being quantized). As such, the length-$2L$ input vectors used for training CSID are given by
\begin{align}
	\mathbf{x}_m & = \begin{bmatrix} Q_{I_1}\left(\Re(x[m])\right) \\ Q_{I_1}\left(\Im(x[m])\right) \\ \vdots \\ Q_{I_{L}}\left(\Re(x[m{-}L{-}1])\right) \\ Q_{I_{L}}\left(\Im(x[m{-}L{-}1])\right) \end{bmatrix}^T, \quad \forall m \in \mathcal{M}_{\text{train}},
\end{align}
where $\Re(x)$ and $\Im(x)$ denote the real and imaginary part of $x$, respectively, and $Q_{I_l}(x),~l \in \{1,\hdots,L\},$ denotes an arbitrary quantization function that maps the values of $x$ to the set $\{1,\hdots,I_l\},~l \in \{1,\hdots,L\}$.

After training is completed, each non-linear SI cancellation sample $\hat{y}_{\text{SI,CSID}}[n]$ is obtained as
\begin{align}
	\hat{y}_{\text{SI,CSID}}[n]	& = \mathcal{X}\left(i_1, \hdots, i_{2L}\right), \label{eq:tensorsi}
\end{align}
where the indices are given by $i_{2l{-}1} = Q_{I_l}(\Re(x[n{-}l{-}1]))$ and $i_{2l} = Q_{I_l}(\Im(x[n{-}l{-}1])),~l \in \{1,\hdots,L\}$, and \eqref{eq:tensorsi} is evaluated efficiently using~\eqref{eq:tensor}.

\subsection{Complexity}

\subsubsection{Memory Complexity}
Since CSID-based SI cancellation uses a complex-valued tensor $\mathcal{X}$, two memory positions are required to store the real and imaginary parts of each complex-valued entry of the factor matrices. Thus, the storage of each matrix $\mathbf{A}_{n}$ requires $2I_nF$ memory locations. As there is a total of $2L$ such matrices and if we also include the $2L$ memory locations required to store the parameters of the linear canceller, the total number of memory locations is
\begin{equation}
     N_{\text{mem,CSID}}= 2\left(F\sum_{n=1}^{2L}I_{\left\lceil\frac{n}{2}\right\rceil}+L\right). \label{eq:memory}
\end{equation}

\subsubsection{Computational Complexity}
CSID-based SI cancellation operates on complex numbers while other methods, such as the polynomial cancellers and most of the existing NN-based methods, operate on real numbers. To make a fair omparison, we compare the number of real-valued multiplications and real-valued additions for each method. Let $a,b \in \mathbb{C}$ and let $a_R = \Re(a)$ and $a_I = \Im(a)$. Complex addition is calculated as 
\begin{align}
a+b & =(a_R+b_R)+j(a_I+b_I ),
\end{align}
and thus requires two real additions. Complex multiplication, on the other hand, can be written in the following two equivalent ways
\begin{align}
ab	& = (a_Rb_R{-}a_Ib_I) + j(a_Rb_I{+}a_Ib_R) \label{eq:cmstandard} \\
	& = (a_Rb_R{-}a_Ib_I) + j((a_R{+}a_I)(b_R{+}b_I){-}a_Rb_R{-}a_Ib_I) \label{eq:cmlowmult}
\end{align}
The expression in~\eqref{eq:cmstandard} requires two real additions and four real multiplications, while the expression in~\eqref{eq:cmlowmult} requires five real additions and three real multiplications. Since multiplications typically have higher implementation complexity, we assume that complex multiplications are implemented using~\eqref{eq:cmlowmult}. As can be seen in~\eqref{eq:tensor}, CSID uses $(F-1)$ complex additions and $F(N-1)$ complex multiplications to calculate one element of $\mathcal{X}$. Moreover, the linear canceller requires a total of $7L-2$ real additions and $3L$ real multiplications. As such, the total number of real additions is
\begin{align}
    N_{\text{add,CSID}} & = F(10L-3)+7L-4, \label{eq:additions}
\end{align}
while the total number of real multiplications is
\begin{align}
   N_{\text{mult,CSID}} & = (6F+1)L-3. \label{eq:multiplications}
\end{align}
We note that, even though the number of quantization levels that is used for each input does not affect the computational complexity of the inference step, it does affect the computational complexity of the training step.

\section{Results}\label{sec : results}

In this section we explore the SI cancellation performance, memory usage, and computational complexity of the CSID-based canceller described in Section~\ref{sec : method}. Moreover, we compare the CSID-based canceller with polynomial and NN-based cancellers in terms of their complexity at a similar SI cancellation performance.

\begin{figure}[t]
\centerline{\begin{tikzpicture}

	\pgfplotsset{grid style={dashed}}

	\begin{axis}[
		width = 0.95\columnwidth,
		height = 0.7\columnwidth,
		xlabel = {Quantization bit-width $\log_2 I$ (bits)},
		ylabel = {Non-linear SIC (dB)},
		ylabel near ticks,
		xlabel near ticks,
		%xtick distance=1,
		ytick distance=2,
		label style={font=\small},
		tick label style={font=\footnotesize},
		xmin = 2, xmax = 7,
		ymin = 0, ymax = 18,
		ymajorgrids,
		legend style={at={(0.16,0.96)},anchor=north,font=\footnotesize},
		legend columns=1,
		legend cell align=left,		
		legend entries={$F=1$,
						$F=2$,
						$F=3$,
						$F=4$,
						$F=5$,
		}                    
	]

		\addplot[Set1-7-1, thick, dotted, mark=*, mark options={solid, ultra thick}] table[x index = 0, y index = 1] {fig/data/results_CPD_L2_canc.dat};
		\addplot[Set1-7-2, thick, dotted, mark=square*, mark options={solid, ultra thick}] table[x index = 0, y index = 2] {fig/data/results_CPD_L2_canc.dat};
		\addplot[Set1-7-3, thick, dotted, mark=diamond*, mark options={solid, ultra thick, scale=1.25}] table[x index = 0, y index = 3] {fig/data/results_CPD_L2_canc.dat};
		\addplot[Set1-7-4, thick, dotted, mark=triangle*, mark options={solid, ultra thick, scale=1.25}] table[x index = 0, y index = 4] {fig/data/results_CPD_L2_canc.dat};
		\addplot[Set1-7-5, thick, dotted, mark=star, mark options={solid, ultra thick, scale=1.4}] table[x index = 0, y index = 5] {fig/data/results_CPD_L2_canc.dat};
		%\addplot[Set1-7-6, thick, dotted, mark=square*, mark options={solid, ultra thick}] table[x index = 0, y index = 6] {fig/data/results_CPD_L2_canc.dat};
		%\addplot[Set1-7-7, thick, dotted, mark=square*, mark options={solid, ultra thick}] table[x index = 0, y index = 7] {fig/data/results_CPD_L2_canc.dat};

	\end{axis}

\end{tikzpicture}%
}
\caption{Achievable non-linear SI cancellation as a function of the quantization bit-width ($\log_2 I$) for various tensor ranks $F \in \{1,\hdots,5\}$.}
\label{fig:quantcanc}
\end{figure}
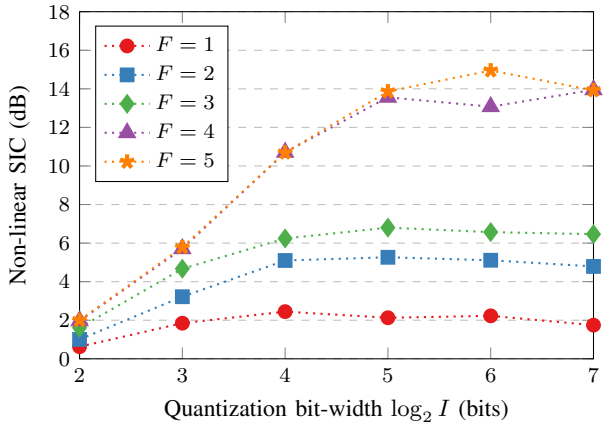

\subsection{Dataset \& Comparison Setup}
We use the dataset of~\cite{Kurzo2020}, where the transmitted baseband signal $x[n]$ is a $20$~MHz QPSK-modulated OFDM signal with $2048$ carriers and a peak-to-average power ratio (PAPR) of $13$~dB. This signal is transmitted over a FD testbed implemented using the National Instruments PXI platform~\cite{NIPXI} in conjuction with a Skyworks SE2576L PA~\cite{SkyworksSE2576L} which operates at its $1$~dB compression point, namely at an  output power of approximately $32$~dBm. We use an RF carrier frequency of $2.45$~GHz and we set the sampling rate of the receiver to $80$~MHz so that we oversample the OFDM signal by a factor of $4$. The dataset contains $20\,480$ time-domain SI baseband samples, of which 80\% are used for training, 10\% for validation, and 10\% as a test set for the final evaluation of the SI cancellation performance.

To implement the CSID-based SI canceller, we modified the MATLAB code provided by the authors of~\cite{Kargas2020}.\footnote{The original MATLAB code is available at~\url{https://github.com/nkargas/Canonical-System-Identification}, while our modified MATLAB and the FD dataset are available at~\url{https://github.com/abalatsoukas/CSI-full-duplex}. Both implementations use the Tensor Toolbox for MATLAB~\cite{TTB_Sparse}.} K-means clustering is used for the quantization and we assume $I_1 = \hdots = I_L = I$ for simplicity. Moreover, we also assume $\mu_1 = \hdots = \mu_{2L} = \mu$ for simplicity. Similarly to the NN-based canceller in~\cite{Kurzo2020}, we use $L=2$ for the CSID-based canceller.

\subsection{SI Cancellation Performance}\label{sec:cancperf}

In order to evaluate the SI cancellation performance of the CSID-based canceller, for every combination of $F$ and $I$, we perform training for $\mu \in \{10^{-6}, 10^{-5}, 10^{-4}, 10^{-3}\}$ and $\rho \in \{10^{-4}, 10^{-3}, 10^{-2}, 10^{-1}\}$ for regularization and smoothing, respectively. The best combination of $(\mu,\rho)$ is selected based on the validation set and the final SI cancellation results are based on the test set.

In Fig.~\ref{fig:quantcanc}, we show the non-linear SI cancellation performance of the CSID-based canceller for $F \in \{1,\hdots,5\}$ and $I \in \{4,8,16,32,64,128\}$. We observe that, as expected, increasing $F$ results in better SI cancellation performance. However, there is a diminishing returns effect after $F=4$ for this particular dataset. Similarly, increasing the quantization bit-width generally improves the SI cancellation performance. The best non-linear SI cancellation is $14.9$~dB and is achieved by using a combination of $F=5$ and $I = 64$.

\subsection{Complexity}
The computational complexity of the CSID-based canceller only depends on the number of inputs $N$ and on the number of factors $F$, as can be seen in~\eqref{eq:additions} and \eqref{eq:multiplications}. Since $N=2L = 4$ is fixed in our experiment, for every value of $F \in \{1,\hdots,5\}$ we perform training for $I \in \{4,8,16,32,64,128\}$, as well as $\mu \in \{10^{-6}, 10^{-5}, 10^{-4}, 10^{-3}\}$ and $\rho \in \{10^{-4}, 10^{-3}, 10^{-2}, 10^{-1}\}$ for regularization and smoothing, respectively. We then select the combination of $(I,\mu,\rho)$ that results in the best SI cancellation performance on the validation set and we report the final performance based on the test set. The results of this analysis are summarized in Fig.~\ref{fig:sic}.

The memory complexity, on the other hand, depends on the number of inputs $N$, the number of factors $F$,  and the number of quantization levels $I$, as can be seen in~\eqref{eq:memory}. As such, the memory complexity analysis is performed in the same way as the SI cancellation performance analysis in Section~\ref{sec:cancperf} and the results are shown in Fig.~\ref{fig:memcanc}. We observe that, interestingly, the best SI cancellation performance for a given memory complexity is achieved by different values of $F$ depending on the operating regime.

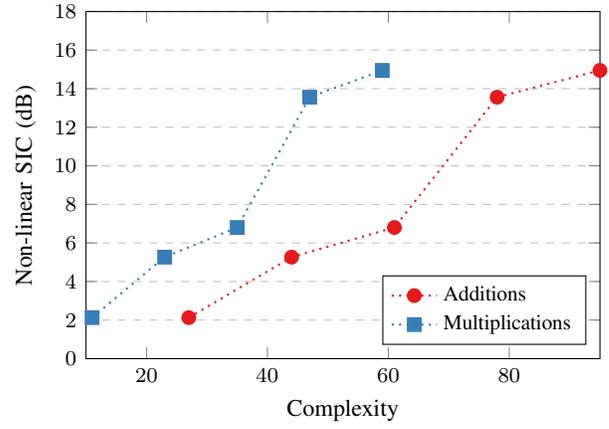
\begin{figure}[t]
\centerline{\begin{tikzpicture}

	\pgfplotsset{grid style={dashed}}

	\begin{axis}[
		width = 0.95\columnwidth,
		height = 0.7\columnwidth,
		xlabel = {Complexity},
		ylabel = {Non-linear SIC (dB)},
		ylabel near ticks,
		xlabel near ticks,
		%xtick distance=1,
		ytick distance=2,
		label style={font=\small},
		tick label style={font=\footnotesize},
		xmin = 10, xmax = 95,
		ymin = 0, ymax = 18,
		ymajorgrids,
		legend style={at={(0.775,0.24)},anchor=north,font=\footnotesize},
		legend columns=1,
		legend cell align=left,		
		legend entries={Additions,
                    Multiplications,
		}                    
	]

		\addplot[Set1-7-1, thick, dotted, mark=*, mark options={solid, ultra thick}] table[x index = 0, y index = 1] {fig/data/results_CPD_L2_adds.dat};
		\addplot[Set1-7-2, thick, dotted, mark=square*, mark options={solid, ultra thick}] table[x index = 0, y index = 1] {fig/data/results_CPD_L2_mults.dat};

	\end{axis}

\end{tikzpicture}%
}
\caption{Achievable non-linear SI cancellation as a function of the computational complexity per sample.}
\label{fig:sic}
\end{figure}
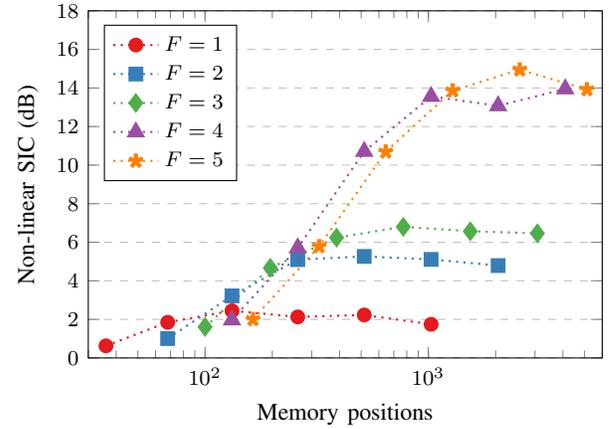
\begin{figure}[t]
\centerline{\begin{tikzpicture}

	\pgfplotsset{grid style={dashed}}

	\begin{semilogxaxis}[
		width = 0.95\columnwidth,
		height = 0.7\columnwidth,
		xlabel = {Memory positions},
		ylabel = {Non-linear SIC (dB)},
		ylabel near ticks,
		xlabel near ticks,
		%xtick distance=1,
		ytick distance=2,
		label style={font=\small},
		tick label style={font=\footnotesize},
		xmin = 30, xmax = 6e3,
		ymin = 0, ymax = 18,
		ymajorgrids,
		legend style={at={(0.16,0.96)},anchor=north,font=\footnotesize},
		legend columns=1,
		legend cell align=left,		
		legend entries={$F=1$,
						$F=2$,
						$F=3$,
						$F=4$,
						$F=5$,
		}                    
	]

		\addplot[Set1-7-1, thick, dotted, mark=*, mark options={solid, ultra thick}] table[x index = 0, y index = 1] {fig/data/results_CPD_L2_mem.dat};
		\addplot[Set1-7-2, thick, dotted, mark=square*, mark options={solid, ultra thick}] table[x index = 2, y index = 3] {fig/data/results_CPD_L2_mem.dat};
		\addplot[Set1-7-3, thick, dotted, mark=diamond*, mark options={solid, ultra thick, scale=1.25}] table[x index = 4, y index = 5] {fig/data/results_CPD_L2_mem.dat};
		\addplot[Set1-7-4, thick, dotted, mark=triangle*, mark options={solid, ultra thick, scale=1.25}] table[x index = 6, y index = 7] {fig/data/results_CPD_L2_mem.dat};
		\addplot[Set1-7-5, thick, dotted, mark=star, mark options={solid, ultra thick, scale=1.4}] table[x index = 8, y index = 9] {fig/data/results_CPD_L2_mem.dat};
		%\addplot[Set1-7-6, thick, dotted, mark=square*, mark options={solid, ultra thick}] table[x index = 0, y index = 6] {fig/data/results_CPD_L2_mem.dat};
		%\addplot[Set1-7-7, thick, dotted, mark=square*, mark options={solid, ultra thick}] table[x index = 0, y index = 7] {fig/data/results_CPD_L2_mem.dat};

	\end{semilogxaxis}

\end{tikzpicture}%
}
\caption{Achievable non-linear SI cancellation as a function of the required memory positions for various tensor ranks $F \in \{1,\hdots,5\}$.}
\label{fig:memcanc}
\end{figure}

\subsection{Comparison with Polynomial and NN-Based SI Cancellers}
For the polynomial canceller, we select $L=3$ and $P=7$. This combination was shown in~\cite{Kurzo2020} to lead to the smallest number of multiplications while maintaining a SI cancellation performance at most 1 dB lower than the maximum achievable SI cancellation performance over a wide range of values for $L$ and $P$ and for the same dataset that we use in this work. We use the \emph{equi-performance} NN-based canceller from~\cite{Kurzo2020}, which uses $L=2$ and $N_h=8$ hidden neurons and has SI cancellation performance that is as close as possible to the polynomial canceller with $L=3$ and $P=7$. Finally, we select the CSID-based canceller with $F=4$ and $I=32$, whose SI cancellation performance is slightly better than both the polynomial and the NN-based canceller.

The aforementioned SI cancellers are summarized in Table~\ref{tab:perf}. We observe that the CSID-based SI canceller requires $81$\% and $5$\% fewer additions than the polynomial and the NN-based cancellers, respectively. More importantly, the CSID-based SI canceller also requires $74$\% and $22$\% fewer multiplications than the polynomial and the NN-based cancellers, respectively. It can also be seen that the CSID-based canceller requires significantly more memory than both the polynomial and the NN-based cancellers.

\section{Conclusion}
In this work, we showed that CSID can be used to effectively model and cancel the non-linear SI signal in FD radios. In particular, we showed that, for the dataset of~\cite{Kurzo2020}, up to 14.9 dB of non-linear SI cancellation can be achieved when using $F=5$ factors and $I=64$ quantization levels. Moreover, we compared the CSID-based SI canceller to the polynomial and the NN-based SI cancellers of~\cite{Kurzo2020} and we showed that, for the same SI cancellation performance, CSID-based cancellation has a significantly lower computational complexity in terms of the required number of multiplications and additions. 
\begin{table}[t]
  \centering
  \caption{Comparison of the non-linear SI cancellation performance and the complexity of the polynomial canceller, the NN-based canceller, and the CSID-based canceller.}\label{tab:perf}
  \begin{tabular}{lrrr}
  \toprule
                          & \multicolumn{1}{c}{Poly.}  & \multicolumn{1}{c}{NN}      & \multicolumn{1}{c}{CSID} \\
    \midrule
    Canc. (dB)     & $11.5$ 	& $13.3$  	& $13.6$ \\
    \midrule
    $L$                   & $3$  	& $2$		& $2$ \\
    $P$                   & $7$  	& n/a		& n/a \\
    $N_h$                 & n/a 	& $8$		& n/a \\
	$F$					  & n/a		& n/a		& $4$ \\
	$I$					  & n/a		& n/a		& $32$ \\
    \midrule
	Additions     	 	  & $418$	& $82$   	& $78$ \\
    Multiplications  	  & $180$ 	& $60$   	& $47$ \\
	Memory 				  & $120$	& $58$		& $1028$ \\
    \bottomrule
  \end{tabular}
\end{table}
\bibliographystyle{IEEEtran}
\bibliography{bibliography}

\end{document}